# Physical properties of the noncentrosymmetric superconductor $Mg_{10}Ir_{19}B_{16}$

T. Klimczuk[1,2], F. Ronning[1], V. Sidorov[1,3], R. J. Cava[4] and J.D. Thompson[1]

[1]Condensed Matter and Thermal Physics, Los Alamos National Laboratory,

Los Alamos, NM 87545, USA

[2]Faculty of Applied Physics, Gdansk University of Technology, Narutowicza 11/12,

80-952 Gdansk, Poland

[3]Vereshchagin Institute for High Pressure Physics, Russian Academy of Sciences, 142190

Troitsk, Russia

[4]Department of Chemistry, Princeton University, Princeton NJ 08544, USA

**Abstract**

Specific heat, electrical resistivity and magnetic susceptibility measurements on a high quality sample of $Mg_{10}Ir_{19}B_{16}$ provide a self-consistent determination of its superconducting properties. They indicate that $Mg_{10}Ir_{19}B_{16}$ is a type–II superconductor ($T_C$=4.45 K, $\kappa(0) \approx 20$), with an electron-phonon coupling constant $\lambda_{ep}$=0.66. An analysis of the T-dependent specific heat shows that superconducting properties are dominated by an s-wave gap ($\Delta$=0.7 meV). Point contact tunneling data provides evidence for multiple superconducting gaps, as expected from strong asymmetric spin-orbit coupling.

The discovery of superconductivity in CePt$_3$Si [1] has stimulated theoretical and experimental efforts to understand the role of the lack of crystal-structure inversion symmetry on superconductivity. Non-centrosymmetric heavy fermion superconductors [1,2,3,4] have attracted the most attention due to the observed unconventional behaviors in these strongly correlated electron compounds. However, transition-metal compounds such as Li$_2$M$_3$B (M = Pd, and Pt)[5,6], Mg$_{10}$Ir$_{19}$B$_{16}$ [7], and M$_2$Ga$_9$ (M=Rh, Ir) [8], are more straightforward for exploring the basic effects derived from breaking inversion symmetry. The asymmetric spin-orbit coupling (SOC) in non-centrosymmetric compounds leads to breaking parity conservation and, therefore, the strength of the SOC, which is determined by the crystallographic structure and elemental composition, has a nontrivial effect on the symmetry of Cooper pairs [9,10,11]. NMR [12] and magnetic penetration depth [13] measurements on Li$_2$M$_3$B highlight the role of composition: increasing SOC by replacing Pd with Pt changes the superconducting order parameter from dominantly spin-singlet (Li$_2$Pd$_3$B) to nodal, spin-triplet (Li$_2$Pt$_3$B).

Mg$_{10}$Ir$_{19}$B$_{16}$ crystallizes in a large (a=10.5668Å, I-43m), rather complex structure whose atomic positions are given in reference 7 and will be discussed elsewhere [14] in more detail. The non-centrosymmetry in Mg$_{10}$Ir$_{19}$B$_{16}$ is global, coming from Ir3 (24g site), Mg1 (8c site), and both B atoms (8c and 24g sites). Because Ir is a heavy transition metal, SOC is expected to have a significant influence on the properties, as Pt does in Li$_2$Pt$_3$B. We find that the thermodynamics of Mg$_{10}$Ir$_{19}$B$_{16}$ can be satisfactorily described with a conventional s-wave gap, while tunneling data present evidence for multiple gaps.

Mg$_{10}$Ir$_{19}$B$_{16}$ samples were synthesized by standard solid state reaction of pure Mg, Ir and B elements as described in ref. [7], with one important exception. The last heating was performed in a sealed Ta tube for 100 hours at 850°C. The sample purity was confirmed by powder X-ray diffraction using Cu Kα radiation on a diffractometer; no extraneous second phases could be detected at the 5% level. Synthesis in a Ta tube allows heating the material for relatively long periods, which drastically improves chemical purity in comparison to the previous method [7].

Measurements of the DC magnetic susceptibility in a SQUID magnetometer (MPMS Quantum Design) showed that $Mg_{10}Ir_{19}B_{16}$ exhibits a temperature-independent Pauli susceptibility above ~ 100K, below which there is a Curie-like tail that increases $\chi(T)$ by about 30% at 2K.[15] The temperature dependence of the zero field cooled (ZFC) and field cooled (FC) magnetic susceptibility in a field of 10 Oe is plotted in the inset of Fig.1a. The superconducting transition temperature found in these measurements is $T_{c\ onset}$ = 4.4 K, which is close to the previous report [7]. At 2K, the diamagnetic response, after demagnetization correction, is about $0.95(1/4\pi)$. The much lower FC signal is an indication of a substantial pinning effect, possibly on grain boundaries, that is reflected as well in M(H) hysteresis (not shown).

The heat capacity was measured using an adiabatic relaxation calorimeter (Quantum Design PPMS). Figure 1a shows $C/T$ versus $T^2$ in the temperature range from 0.4K to 7.5K in different magnetic fields. The bulk nature of the superconductivity and good quality of the sample is confirmed by a sharp anomaly at $T_c$ = 4.45K which is consistent with $T_{c\ onset}$ determined by $\chi(T)$. In zero field a small but clear residual linear term $\gamma_0$=2.5 mJ/mol K$^2$ is observed in the T→0K limit (see also figure 2 inset). While $\gamma_0$ could originate from a nodal superconducting gap provided that the impurity bandwidth is sufficiently large, we attribute it more likely to a small fraction of non-superconducting impurity phases. The measurement at $\mu_0H$=5T, which exceeds $H_{c2}$, was fitted using the formula $C = \gamma T + \beta T^3 + \delta T^5$ and gives the parameters, $\gamma$=55.1(1) mJ/mol K$^2$, $\beta$=3.97(1) mJ/mol K$^4$ and $\delta$=7.4(2) µJ/mol K$^6$. The first and last two terms are attributed to the electronic and lattice contribution to the heat capacity, respectively. By subtracting the impurity concentration $\gamma_0$ we find the Sommerfeld coefficient for Mg$_{10}$Ir$_{19}$B$_{16}$ $\gamma_n = \gamma - \gamma_0 = 52.6$ mJ/mol K$^2$. Knowing $\gamma_n$ we can calculate the normalized specific heat jump, $\Delta C/\gamma_n T_c = 1.60$, which suggests an enhanced electron – phonon coupling. In a simple Debye model for the phonon contribution, the $\beta$ coefficient is related to the Debye temperature ($\Theta_D$) through $\Theta_D = \left(\dfrac{12\pi^4}{5\beta}nR\right)^{1/3}$, where R=8.314 J/mol K and n=45 for Mg$_{10}$Ir$_{19}$B$_{16}$. However, a $\Theta_D = 280$ K derived from this relation strongly overestimates the heat capacity between 10 and 300 K as shown in the inset of Fig. 1b. This indicates that a simple Debye model that accurately models the acoustic modes at low energies fails to capture the higher energy optical modes that must be present in this material. To estimate the role of optical phonons, we fit the data above 10K to a Debye model plus an Einstein mode, as was done for MgB$_2$.[16] The fit gives 63% of the weight to a Debye term with $\Theta_D = 740$ K, and the remaining weight in an Einstein mode with energy $\Theta_E = 147$ K.

The electron – phonon coupling constant ($\lambda_{ep}$) can be estimated from the modified McMillian formula [17,18]: $\lambda_{ep} = \dfrac{1.04 + \mu^* \ln(\omega_{\log}/1.2T_C)}{(1-0.62\mu^*)\ln(\omega_{\log}/1.2T_C) - 1.04}$ where $\mu^*$ is a Coulumb repulsion constant and $\omega_{\log}$ is a logarithmic averaged phonon frequency and can be determined from $\Delta C/\gamma T_C|_{T_C} = 1.43\left[1 + 53(T_C/\omega_{\log})^2 \ln(\omega_{\log}/3T_C)\right]$. Taking $\mu^* = 0.10$, we determined $\omega_{\log} =$

145K, and $\lambda_{ep}$=0.66. This value suggests that $Mg_{10}Ir_{19}B_{16}$ is a moderate – coupling superconductor[19].

The temperature dependence of the electronic specific heat ($C_{el}$) below $T_c$ is shown in Fig. 1b. Three models were used to fit the data, $C \propto T^2$, $T^3$, and $e^{-b/kT}$ expected for line nodes, point nodes, and a conventional, fully gapped model, respectively. The small residual linear term ($\gamma_0 T$) of 2.5*T (mJ/mol K), was held constant throughout the fits. The best fit is provided by a fully gapped mode. An s-wave BCS model of the entire $C_{es}(0T)$ data gives $2\Delta_0 = 1.4$meV or 16K, comparable to the energy gap in $Li_2Pt_3B$ (7.7K)[20] and $Li_2Pd_3B$ (29.6K)[20].

Dominance of a s-wave channel is supported by measurements of the magnetic field dependence of the Sommerfeld parameter $\gamma(H)$. For a highly anisotropic gap or a gap with nodes, theory predicts a nonlinear $\gamma(H) \propto H^{1/2}$ dependence[21]. In contrast, for a fully gapped superconductor, $\gamma(H)$ should be proportional to the number of field-induced vortices, i.e. $\gamma(H) \propto H$. As shown in Fig. 2, the field-linear increase in $\gamma(H)$ suggests that most electronic states near the Fermi energy are gapped in $Mg_{10}Ir_{19}B_{16}$ and clearly is at odds with a $\gamma(H) \propto H^{1/2}$ dependence.

In Fig. 3, the resistively determined upper critical field (Fig. 3 inset) and that obtained from specific heat measurements (see Fig. 1) coincide within experimental uncertainty at low fields but diverge with increasing field. As the magnetic field is raised, the resistive transition width increases from 0.1K to 0.8K for fields of 0T and 1.2T, respectively. One interpretation of the transition broadening is that it arises from filamentary-like superconductivity along grain boundaries, where scattering is stronger. The associated reduced electronic mean free path, in turn, would decrease the intrinsic coherence length and raise the resistive $H_{c2}$ relative to the bulk $H_{c2}$ determined by specific heat, which we take to be intrinsic. As shown by the dashed line in Fig. 3, the temperature dependence of $H_{c2}(T)$ is described by the Werthamer - Helfand – Hohenberg (WHH) [22] expression for a dirty type-II superconductor and gives $\mu_0 H_{c2}^{WHH}(0) = 0.77(2) T$, consistent with results in Fig. 3. This value of $H_{c2}(0)$ for $Mg_{10}Ir_{19}B_{16}$ is smaller than previously reported[7]; however, the purity of the present sample is much better, resulting in a longer mean-path, and the earlier result was based only on resistivity measurements which we have found can lead to a much higher $H_{c2}(0)$. Assuming a Lande g-factor of 2, the measured $H_{c2}(0)$ is well below the weak-coupling Pauli field of $\approx 8.2T$, indicating that the observed critical field is dominated by orbital pair-breaking.

The lower critical field values, $H_{c1}(T)$, were determined from low-field M(H) curves in which special care was taken to correct for a small residual trapped field in the superconducting magnet and for demagnetization effects. With these precautions, $H_{c1}(T)$ was defined as the field at which M(H) deviated by 1% from a perfect diamagnetic response. Values are plotted in the inset of Fig. 3, and their extrapolation to T=0 K gives $\mu_0 H_{c1}(0) = 3mT$.

With these results for $H_{c1}(0)$ and $H_{c2}(0)$, we can estimate several superconducting parameters for $Mg_{10}Ir_{19}B_{16}$. From $\mu_0 H_{c2} = \frac{\Phi_0}{2\pi \xi_{GL}^2}$, where $\Phi_0$ is the quantum flux (h/2e), we find a Ginzburg-Landau coherence length $\xi_{GL}(0) = 206$ Å. Knowing $\xi_{GL}(0)$ and $H_{c1}(0)$, a penetration depth, $\lambda_{GL}(0) = 4040$ Å, is obtained from $\mu_0 H_{c1} = \frac{\Phi_0}{4\pi \lambda_{GL}^2} \ln \frac{\lambda_{GL}}{\xi_{GL}}$, and, hence, the Ginzburg – Landau parameter, $\kappa(0) \approx 20$. Using these parameters and the relation $H_{c1} \cdot H_{c2} = H_c^2 \ln(\kappa)$, we find that the thermodynamic critical field $\mu_0 H_c(0)=28mT$. The superconducting condensation energy provides a stringent self consistency check on the derived parameters. The condensation energy relates the thermodynamic critical field to measured specific heat difference between zero field and the normal state through: $\mu_0 H_c^2(0)/2 = F_N - F_S = \iint (C(5T) - C(0T))/T dT$. This method of calculating the thermodynamic critical field gives $\mu_0 H_c(0)=30$mT, in excellent agreement with $\mu_0 H_c(0)$ determined by the critical fields. In addition the condensation energy is related to the density of states and the gap value by $\mu_0 H_c^2(0) = N(E_F)(1+\lambda_{ep})\Delta_0^2$. We obtain $N(E_F)(1+\lambda_{ep})$ from $\gamma_n$ and calculate $\Delta_0$ to be 0.62 meV, which agrees very well with the value of 0.7 meV obtained directly from a fit to the zero field specific heat.

While a single s-wave gap is indicated by transport and thermodynamic measurements, we obtain a different view from point contact spectroscopy measurements[23]. A representative normalized conductance curve is shown in figure 4. Multiple energy scales are observed as marked by the arrows. Tracking the temperature dependence of these energy scales (inset of fig. 4) shows that both correlate well with the BCS gap expectation, and it also shows that we are not in the thermal regime which would give a $(1-(T/T_c)^2)^{1/2}$ dependence. Furthermore, a simple fit assuming pure Andreev reflection from a multiband superconductor with two gaps (solid line) gives values of 0.56meV and 2.2meV. We note that the more dominant lower energy scale is in reasonable agreement with the gap obtained by our thermodynamic measurements, but that a single gap fit (dashed line) giving $\Delta = 0.84$meV can not capture the shoulder above 1meV.

The observation of multiple gaps is as expected, due to the mixing of the spin singlet and spin triplet order parameters. However, no conclusive evidence for such physics was observed by our other measurements. We could speculate that the larger gap would manifest itself most strongly in the specific heat data near $T_c$, but that inhomogeneity may blur the features. Perhaps, due to the local nature of the probe, point contact tunneling can more easily resolve multiple gap features. A more detailed theory is needed to analyze the complete set of data to extract the full multi-gap structure, and measurements on single crystals would be useful in resolving the apparent discrepancy between thermodynamic and tunneling measurements.

In summary, a self-consistent set of superconducting and normal states parameters confirm the validity of measured and derived properties from a high quality sample of $Mg_{10}Ir_{19}B_{16}$ and provide a benchmark for refined band structure calculations that explicitly include the effect of SOC that must be present. Asymmetric spin-orbit coupling arising from the heavy element Ir should spin-split degenerate electronic bands by a factor many times $k_BT_c$, and, therefore, allow mixing of spin-singlet and spin-triplet pairing states [10]. A single band with a gap of 0.7meV can account for the large majority of the data. In contrast, we find clear evidence for multiple gaps from tunneling measurements. Nuclear spin-lattice relaxation and Knight-shift measurements on ideally purer samples should reveal unambiguously the presence or absence of a nodal gap structure and the existence of a spin-triplet component in the superconducting state.

Upon completion of this work we became aware of (ref. 24), which has similar conclusions with our thermodynamic measurements. We thank H. Q. Yuan for useful discussions. Work at Los Alamos and Princeton was performed under the auspices of the US DOE, Office of Science.

**Figure Captions**

**Figure 1.** (Color online) (a) Specific heat divided by temperature (C/T) as a function of $T^2$ under magnetic field from 0T to 1T increasing by steps of $\Delta\mu_0H=0.1T$. The inset shows the superconducting transition measured in zero-field-cooled (ZFC) and field-cooled (FC) mode. (b) Temperature dependent electronic contribution to the specific heat of $Mg_{10}Ir_{19}B_{16}$. Lines are fits described in the text. The inset shows lattice specific heat ($C_{ph}$) as a function of temperature from 2 to 300K. The blue dotted line represents a Debye model expectation ($\Theta_D = 280$ K) and black solid line is a fit to a Debye model ($\Theta_D = 740$ K) plus an Einstein mode fit ($\Theta_D = 147$ K). See text for details.

**Figure 2.** (Color online) Magnetic field dependence of $\gamma(H)$ as a function of magnetic field $\mu_0H$. The solid (black) line corresponds to a linear (~H) relation and dotted (blue) line represents the non-linear (~$H^{1/2}$) relation. The inset shows C/T vs $T^2$ under magnetic field (from 0T to 0.6T) in the lowest temperature region.

**Figure 3.** (Color online) The upper critical field ($\mu_0H_{c2}$) from specific heat (red squares) and from resistivity (black circles) as a function of temperature. The vertical bars show 5% and 95% of the resistively determined superconducting transition. The dashed curve is predicted by the WHH expression. The upper right inset is the temperature dependence of the lower critical field ($\mu_0H_{c1}$) for $Mg_{10}Ir_{19}B_{16}$. The lower left inset shows the resistivity near $T_c$ for representative applied fields. From the initial slope $-dH_{c2}/dT_c = 2700$ Oe/K (using specific heat data), $\gamma = 52.6$ mJ/molK$^2$, and the relation $\left.\frac{dH_{C2}}{dT}\right|_{T_C} = 4.48\cdot10^4\,\gamma\rho$ (Oe/K), (with $\gamma$ in (erg/cm$^3$K$^2$) and $\rho$ in ($\Omega$cm)), we estimate the intrinsic intragranular resistivity $\rho(T\geq T_C) \approx 80$ $\mu\Omega$ cm. This intragranular resistivity is only about 14% of the value seen in the inset, consistent with strong intergranular scattering and an associated larger $H_{c2}$ measured resistively.

**Figure 4.** (Color online) Normalized conductance spectra at 2.405K (circles) and 4.825K > $T_c$ (thin black line). Arrows indicate the presence of two energy scales in the superconducting state. The dashed (solid) line is a fit assuming pure Andreev reflection with one (two) gap(s). Relaxing this assumption within the Blonder-Tinkham-Klapwijk formalism did not improve the fits. The inset shows the temperature dependence of the two gap scales compared with theoretical expectations.

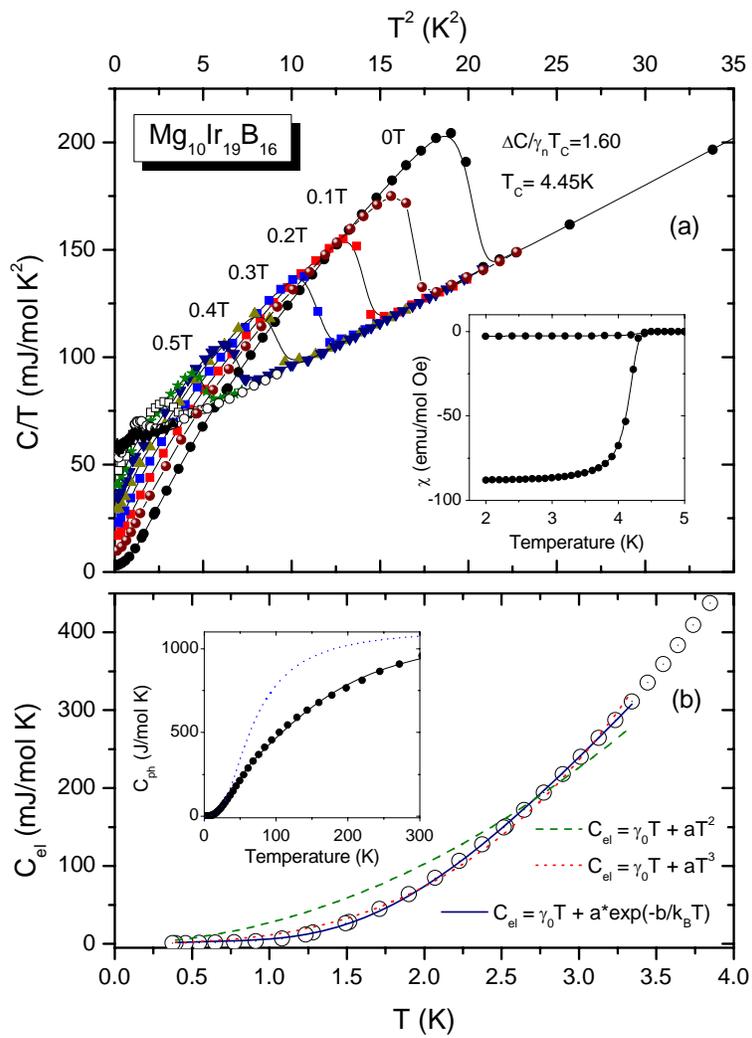

**Fig. 1**

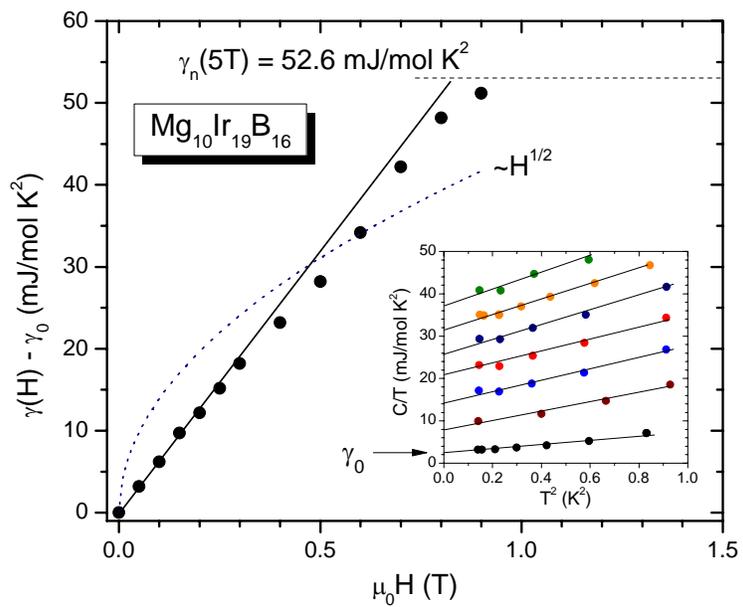

**Fig. 2**

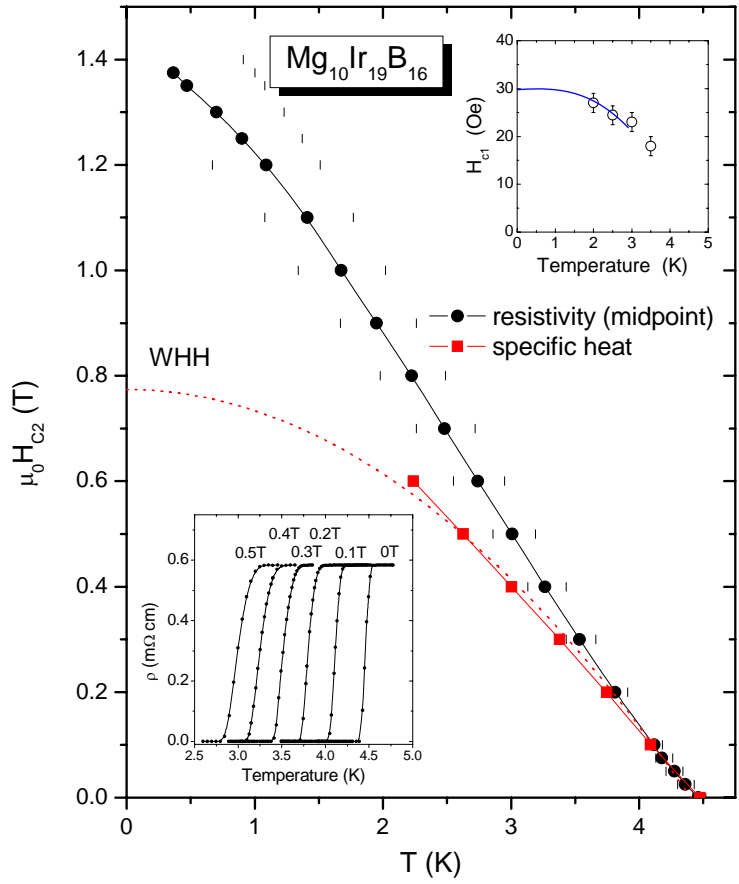

**Fig. 3**

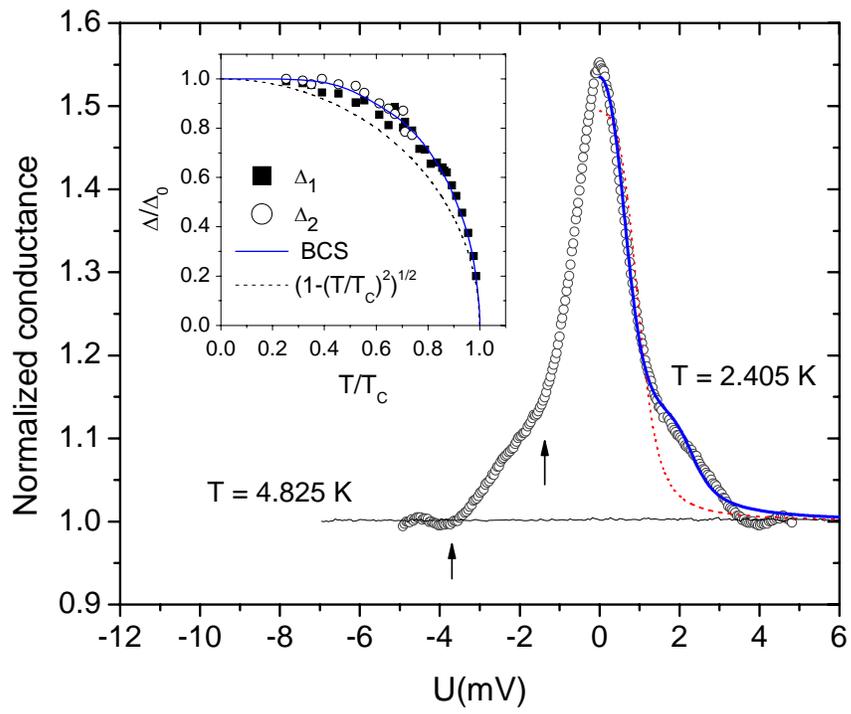

**Fig. 4**

[15] The measured, temperature-independent susceptibility above ~ 100 K is weakly diamagnetic, $\chi = -4.5e^{-4}$ emu/mol. Correcting the measured $\chi$ for core diamagnetic contributions ($\chi_c = -7e^{-4}$ emu/mol) gives a Pauli susceptibility of $2.5e^{-4}$ emu/mol. Assuming a Wilson ratio of 1 and taking values of the Sommerfeld coefficient and electron-phonon coupling parameter, discussed in the text, the calculated Pauli susceptibility is $4.4e^{-4}$ emu/mol, in reasonable agreement with the estimate above.